\documentclass[10pt]{article}
\usepackage{amsmath}
\usepackage{amssymb} 
\newcommand{\R}{\mathbb{R}} % Real numbers
\newcommand{\half}{\frac{1}{2}}

\newcommand{\gcycl}{\operatorname{gcycl}}

\newcommand{\BMf}{\mathcal{M}}

\newcommand{\Poisson}{\mathcal{P}}

\newcommand{\casimir}{c}

\newcommand{\Action}{L}

\newcommand{\N}{N}

\newcommand{\rvec}[1]{\overset{\,
    \raisebox{-0.5ex}{$\scriptscriptstyle \rightarrow$}}{#1}{}}

\newcommand{\rpartial}{\rvec{\partial}}
\newcommand\ap[3]    
		{{\it Ann.\ Phys.\ (NY) }{\bf #1} (#2) #3}
\newcommand\cqg[3]  
		{{\it Class.\ and Quant.\ Grav.\ }{\bf #1} (#2) #3}
\newcommand\ijmpa[3] 
		{{\it Int.\ J.\ Mod.\ Phys.\ }{\bf A #1} (#2) #3}
\newcommand\jhep[3]  
		{{\it J. High Energy Phys.\ }{\bf #1} (#2) #3}
\newcommand\jpha[3]  
		{{\it J. Phys.\ }{\bf A #1} (#2) #3}
\newcommand\jmp[3]   
		{{\it J.\ Math.\ Phys.\ }{\bf #1} (#2) #3}
\newcommand\mpla[3]  
		{{\it Mod.\ Phys.\ Lett.\ }{\bf A #1} (#2) #3}
\newcommand\npb[3]   
		{{\it Nucl.\ Phys.\ }{\bf B #1} (#2) #3}
\newcommand\plb[3]   
		{{\it Phys.\ Lett.\ }{\bf B #1} (#2) #3}
\newcommand\prd[3]   
		{{\it Phys.\ Rev.\ }{\bf D #1} (#2) #3}
\newcommand\prep[3]  
		{{\it Phys.\ Rept.\ }{\bf #1} (#2) #3}
\newcommand\tmp[3]
		{{\it Theor.\ Math.\ Phys.\ }{\bf #1} (#2) #3}

\begin{document}

\begin{titlepage} 
\renewcommand{\thefootnote}{\fnsymbol{footnote}}

\hfill{}TUW--01--24 \\

{\par\centering \vspace{1cm}\par}

{\par\centering \textbf{\Large 
All minimal supergravity extensions\\
 of 2d dilaton theories$^{+)}$ 
}\large \par}

{\par\centering \vspace{2.0cm} %% \vfill
\renewcommand{\baselinestretch}{1}\par}

{\par\centering \textbf{W.\ Kummer${}^{1}$\footnotemark[1] 
M.\ Ertl${}^{1}$\footnotemark[2] and T.\ Strobl${}^{2}$
\footnotemark[3]{}} 
\vspace{7ex}\par}

{\par\centering \( ^{1} \)Institute for Theoretical Physics, 
Vienna University of Technology, Vienna, Austria \vspace{2ex}
\par}

{\par
\centering \( ^{2} \)Institute for Theoretical Physics, 
FSU Jena,  Jena, Germany
\par}
\vspace{10ex}
{\par\centering
\footnotetext[1]{e-mail: \texttt{wkummer@tph.tuwien.ac.at}} 
\footnotetext[2]{e-mail: 
\texttt{ertl@tph.tuwien.ac.at}} \footnotetext[3]{
e-mail:\texttt{pth@tpi.uni-jena.de}}
\par}

\abstract{The formulation of 2d-dilaton theories, like 
spherically reduced Einstein gravity, is greatly facilitated 
in a formulation as a first order theory with nonvanishing 
bosonic torsion. This is especially also true at the 
quantum level. The interpretation of superextensions as 
graded Poisson sigma models is found to cover generically 
all possible 2d supergravities. Superfields and thus 
superfluous auxiliary fields are avoided altogether. The 
procedure shows that generalizations of bosonic 2d models 
are highly ambiguous. }
\\[2.0cm]
\footnoterule
\mbox{}
{\par
\noindent
${}^{+)}\;$ Talk by W.\ Kummer at Int.\ Europhysics Conference for High 
Energy Physics, Budapest, 12-18 July 2001. 
\par}
\end{titlepage}

%% ----------------------------------------

Despite the fact that so far no tangible direct evidence for
supersymmetry has been discovered in nature, supersymmetry
\cite{Wess:1977fn} managed to retain continual interest within the aim to
arrive at a fundamental `theory of everything' ever since its
discovery: first in supergravity \cite{Freedman:1976xh} in $d = 4$, then in
generalizations to higher dimensions of higher $\N$
\cite{Vafa:1996xn}, and finally incorporated as a low energy limit of
superstrings \cite{Witten:1995ex} or of even more fundamental theories
\cite{Sezgin:1997cj} in 11 dimensions.

Even before the advent of strings and superstrings the importance of
studies in $1+1$ `spacetime' had been emphasized \cite{Howe:1979ia}
in connection with the study of possible superspace formulations. 
To the best of our knowledge, however, to this
day there have been only few attempts  to generalize the supergravity
formulation of (trivial) Einstein-gravity in $d = 2$ to the
consideration of two-dimensional $(1, 1)$ supermanifolds for which the
condition of vanishing (bosonic) torsion is removed 
\cite{Ertl:Diss,EKK}. Only attempts to
formulate theories with higher powers of curvature (at vanishing
torsion) seem to exist \cite{Hindawi:1996fy}. There seem to be only
very few exact solutions of supergravity in $d = 4$ as well
\cite{Aichelburg:1978fn}.

Especially at times when the number of arguments in favour of the
existence of an, as yet undiscovered, fundamental theory 
increases 
\cite{Vafa:1996xn} it may seem appropriate to also exploit --- if
possible --- \emph{all}  generalizations of the
two-dimensional stringy world sheet. Actually, such an undertaking can
be (and indeed is) successful, as suggested by the recent much
improved insight, attained for all (non-supersymmetric)
two-dimensional diffeomorphism invariant theories, including dilaton
theory, and also comprising torsion besides curvature \cite{Katanaev:1986wk} 
in the most general manner. In the absence of
matter-fields (non-geometrical degrees of freedom) all these models
are integrable at the classical level and admit the analysis of all
global solutions \cite{Katanaev:1993fu}. 
 Even many aspects of quantization
of any such theory now seem to be well understood
\cite{Kummer:1997hy}. 
By contrast, in the presence of matter and if singularities like black
holes (BH) occur in such models, integrable solutions are known only for
very few cases. These include interactions with fermions of one
chirality \cite{kummer92} and, if scalar fields are present, only the
dilaton black hole \cite{Mandal:1991tz}, ``chiral'' scalars 
\cite{Strobl:1998ch} and models which have
asymptotical Rindler behaviour \cite{Fabbri:1996bz}.  Therefore, a
supersymmetric extension of the matterless case suggests that the
solvability may carry over, in general. Then, at least part 
of  ``matter'' could be
represented by superpartners of the geometric bosonic field 
variables. At the quantum level for bosonic gravity in two 
dimensions the path integral formalism has proved 
to be invaluable to exactly integrate geometry and to treat 
matter in a consistent perturbation theory \cite{Kummer:1997jj}. 
Also this exact geometrical integration should carry over to 
the supergravity case.\\ 
Within the realm of bosonic two dimensional gravity theories, 
including those with nonvanishing torsion, the reformulation 
as a first order action \cite{Schaller:1994es,Strobl:1999Habil} 
 with auxiliary fields 
$\phi$ and $X^a$ ($ Y =  X^aX_a/2$) 
\begin{equation}
  \Action^\mathrm{FOG} = \int_\BMf \phi d\omega + X_a De^a + \epsilon
  v(\phi,Y) 
  \label{FOG}
\end{equation}
has led to new insights. Indeed $\Action^\mathrm{FOG}$ for 
a potential $v$ quadratic in torsion
\begin{equation}
  \label{vdil}
  v^{\mathrm{dil}}(\phi,Y) = Y Z(\phi) + V(\phi).
\end{equation}
is exatly equivalent \cite{Katanaev:1993fu},\cite{Kummer:1997jj} 
to a generalized dilaton theory 
\begin{equation}
  \label{dil}
  \Action^{\mathrm{dil}} = \int d^2\!x \sqrt{-g}
  \left[
    \frac{\widetilde{R}}{2} \phi - \half Z(\phi) (\partial^n \phi)
    (\partial_n \phi) + V(\phi)
  \right].
\end{equation}
where $\tilde{R}$ represents the torsion free curvature. A 
special case of such dilaton theories is spherically reduced 
Einstein gravity (SRG) in $D$ dimensions 
\begin{equation}
\begin{split}
Z_{SRG} &= - \frac{D-3}{D-2} \; \phi^{-1}\\
V_{SRG} &= - \lambda^2\, \phi^{\frac{D-4}{D-2}}
\end{split}
\label{eq:ddim}
\end{equation}
with the Schwarzschild BH solution. Also the 
so-called dilaton BH \cite{Mandal:1991tz}
 ($D\to \infty$ in (\ref{eq:ddim})) and a
``Poincar\'e gauge'' \cite{Hehl:1995ue}  theory 
 quadratic in curvature 
and torsion \cite{Katanaev:1986wk} and simpler models with $Z=0$, like the 
Jackiw-Teitelboim gravity ($v=-\Lambda\phi$) \cite{jackiw84} are 
covered by (\ref{dil}) and thus also by (\ref{FOG}). \\
Our present approach to obtain the minimal supergravity 
extensions of generic models of type (\ref{FOG}) is based upon the 
concept of the Poisson-Sigma models (PSM)  
\cite{Schaller:1994es,Ikeda:1994aa,Schaller:1994uj}.

Collecting zero form and
one-form fields, respectively, within (\ref{FOG}) as
\begin{equation}
  \label{ident1}
  (X^i) := (\phi,X^a), \qquad
  (A_i) = (dx^m A_{mi}(x)) := (\omega,e_a),
\end{equation}
and after a partial integration, the action (\ref{FOG}) may be
rewritten as
\begin{equation}
  \label{PSM}
  \Action^{\mathrm{PSM}} = \int_\BMf dX^i \wedge A_i + \half
  \Poisson^{ij} A_j \wedge A_i,
\end{equation}
where the matrix $\Poisson^{ij}$ may be read off by direct comparison.
The basic observation in this framework is that this matrix defines a
Poisson bracket on the space spanned by target space 
coordinates $X^i$ of a Sigma Model. In the
present context the related  bracket $\{ X^i , X^j \} := \Poisson^{ij}$ has
the form
\begin{align}
  \{ X^a, \phi \} &= X^b \epsilon_b\^a, \label{LorentzB} \\
  \{ X^a, X^b \} &= v(\phi,Y) \epsilon^{ab}\; .  \label{PB}
\end{align} 
This bracket may be verified to obey the Jacobi identity 
$\{ \{X^a, X^b\}, X^c \} + \mbox{cycl. perm.} = 0$. 
Eq.\ (\ref{LorentzB}) shows that $\phi$ is the generator of
Lorentz transformations (with respect to that bracket) on the target
space $\R^3$. Minimal supersymmetric extensions are obtained 
\cite{Ikeda:1994ab,Izquierdo:1998hg} and \cite{Strobl:1999zz}
 by adding additional anticommuting target 
space coordinates $\chi^\alpha$ and corresponding 
Rarita-Schwinger 1-form fields $\psi_\alpha $  in 
(\ref{ident1}), $X^I = (X^i, 
\chi^\alpha)$ and $A_I = (A_i, \psi_\alpha))$. 

Thus the PSM action (\ref{PSM}) generalizes to
\begin{equation}
  \label{gPSM}
  L^\mathrm{gPSM} = \int_\BMf dX^I A_I + \half \Poisson^{IJ} A_J 
  A_I\, .
\end{equation}
Both $\chi^\alpha$ and $\psi_\alpha$ denote Majorana fields, 
when, as in what follows, $N=1$ supergravity is considered. 
The graded Poisson tensor $\mathcal{P}^{IJ} = (-1)^{IJ+1} 
\mathcal{P}^{JI}$ is again assumed to fulfil a ``graded'' Jacobi 
identity 
\begin{equation}
  J^{IJK} = \Poisson^{IL} \rpartial_L \Poisson^{JK} + \gcycl(IJK) 
  = 0 \, .
  \label{gJacobi} 
\end{equation}
Together with (\ref{gJacobi}) the e.o.m-s with right derivatives 
$\rvec{\partial}_I = \frac{\rpartial}{\partial X^I}$
\begin{gather}
  dX^I + \Poisson^{IJ} A_J = 0, \label{gPSM-eomX} \\
  dA_I + \half (\rpartial_I \Poisson^{JK}) A_K A_J = 0
  \label{gPSM-eomA}
\end{gather}
provide the on-shell symmetries of the action (\ref{gPSM}) 
\begin{equation}
  \label{gPSM-symms}
  \delta X^I = \Poisson^{IJ} \epsilon_J, \qquad
  \delta A_I = -d\epsilon_I - (\rpartial_I \Poisson^{JK}) \epsilon_K 
  A_J\, ,
\end{equation}
which depend on infinitesimal local parameters $\epsilon_I = 
(\epsilon_\phi, \epsilon_a, \epsilon_\alpha)$. The mixed 
components $\mathcal{P}^{\alpha\phi} $ are constructed by 
analogy to $\mathcal{P}^{a\phi}$ in (\ref{LorentzB}) 
with the appropriate 
generator $(-\gamma_5/2)$ of Lorentz transformations in 2d 
spinor space. Then $d\epsilon_\alpha$ in the second set of 
eq.\ (\ref{gPSM-symms}) acquires an additional term casting it into the 
covariant $(D\epsilon )_\alpha$, with covariant derivative 
$D$  appropriate for a supergravity transformation. \\
As the Poisson tensor $\mathcal{P}^{IJ}$ is not of full 
rank, Casimir functions $\mathcal{C} (Y,\phi,\chi^2)$ exist 
which are the solutions of $\left\{ X^I, \mathcal{C} 
\right\} = \mathcal{P}^{IJ} \partial_J \mathcal{C} = 0$ and 
thus correspond to conserved quantities $d \mathcal{C} = 
dX^I\, \partial_I \mathcal{C} = 0$ when (\ref{gPSM-eomX}) is used.   
The bosonic $\mathcal{C}$ in  supergravity is of the form 
\begin{equation}
  \label{Casimir}
  \mathcal{C} = \casimir + \half \chi^2 \casimir_2\, ,
\end{equation}
where $\casimir$ and $\casimir_2$ are functions of $\phi$ 
and $Y$ only. 
However, also 
fermionic Casimir functions may occur (see below). In the 
pure bosonic case ($\chi = \psi = 0$) and for the potential 
(\ref{vdil}) the differential equation for $\mathcal{C}$ allows an 
analytic solution. For instance, $c$ for SRG simply coincides 
(up to a factor) with the ADM mass of the Schwarzschild BH. 
It is interesting, though, that such a conservation law 
continues to exist also in interactions with additional 
matter contributions \cite{Kummer:1995qv}, i.e.\ beyond the range of 
validity of the PSM concept. \\
The determination of all possible minimal supergravities 
\cite{EKS,Ertl:Diss} now reduces to finding the solutions of the Jacobi 
identities (\ref{gJacobi}). In the general ansatz for 
$\mathcal{P}^{IJ}$ 
\begin{align}
\mathcal{P}^{ab} &=  V\, \epsilon^{ab}\, ,\label{pija}\\
\mathcal{P}^{b\phi} &=  X^a\, \epsilon_a{}^b\, ,\label{pijb}\\
\mathcal{P}^{\alpha\phi} &=  -\frac{1}{2} \chi^\beta 
(\gamma_5)_\beta{}^\alpha\, ,\label{pijc} \\
\mathcal{P}^{\alpha b} &=  \chi^\beta (F^b)_\beta{}^\alpha\, 
,\label{pijd} \\
\mathcal{P}^{\alpha\beta} &=  v^{\alpha\beta} + 
\frac{\chi^2}{2} \, v_2^{\alpha\beta} \, ,
\label{pije}
\end{align}
the function 
\begin{equation}
\label{yphi}
V = v (\phi,Y) + \frac{\chi^2}{2} \, v_2 (\phi,Y) 
\end{equation}
contains the original bosonic potential $v$. As explained 
above, eqs.\ (\ref{pijb}) and 
(\ref{pijc}) are fixed by Lorentz invariance. Each one of 
the (symmetric) 
spinor-tensors $v^{\alpha\beta}$ and $v_2^{\alpha,\beta}$ 
in (\ref{pije}) can be further 
expanded into three scalar functions of $Y$ and $\phi$, multiplying 
the symmetric matrices $(\gamma_5)^{\alpha\beta}, X^a(\gamma_a)^{\alpha\beta}, 
X^a (\gamma_5\gamma_a)^{\alpha\beta}$. The quantity 
$(F^b)_\beta{}^\alpha$ is easily seen to depend on another 
set of eight scalar functions. Thus the task to solve the Jacobi 
identities (\ref{gJacobi}), which are differential 
equations, at first 
sight seems to be quite formidable.\\
Fortunately, in the course of our extensive analysis 
\cite{EKS,Ertl:Diss} 
it turned out that by starting from the solutions of the 
Casimir functions, obeying equations like the one sketched 
after eq.\ (\ref{Casimir}), the problem, relating the many unknown 
functions above to the original bosonic potential $v$, may be 
reduced to the solution of \textit{algebraic} equations.\\
We have classified the different cases according to the rank 
of $\mathcal{P}^{IJ}$, when the fermionic degrees of freedom 
are included. The bosonic sub-space is odd-dimensional which 
produced one Casimir function $c$. 
For full ``fermionic rank'', i.e.\ when  
the rank of $v^{\alpha\beta}$ in (\ref{pije} 
is two, the single Casimir function 
(\ref{Casimir}) appears  and the general solution still depends on 
five  
arbitrary (bosonic) functions of $\phi$ and $Y$ beside $v$. 
\\
If the fermionic rank is reduced by one, beside the bosonic 
Casimir function (\ref{Casimir}) a fermionic one exists. It is of the 
generic form 
\begin{equation}
\label{phix}
\mathcal{C}^{(\pm)} = 
\chi^\pm \; \left\vert\, 
\frac{X^{--}}{X^{++}}\,\right\vert^{\pm 1/4}\; 
c_{(\pm)}\, (\phi, X)
%% eq:21
\end{equation}
and owes its Lorentz invariance to the abelian boost 
transformation $\exp (\pm \beta)$ of the light cone 
coordinates $X^{\pm\pm}$, related to $X^a$ and $\exp (\pm 
\beta/2)$ of the chiral spinor components $\chi^{\pm}$. Then 
the general solution of the gPSM algebra 
contains four arbitrary functions beside $v$.\\
For rank zero of the fermionic extension, i.e.\ rank three as in 
the pure bosonic case, in $\mathcal{P}^{IJ}$ beside 
(\ref{Casimir}) 
both fermionic Casimir functions (\ref{phix}) are conserved and 
three functions remain arbitrary for a given bosonic 
potential $v$. \\
This arbitrariness can be understood as well 
by studying  
reparametrizations of the target space, spanned by the $X^I$ 
in the gPSM. Those reparametrizations may generate  new 
models. Therefore, they can be useful to create a more 
general gPSM from a simpler one, although this 
approach is 
difficult to handle if $v$ in (\ref{yphi}) is assumed to be 
the given 
starting point. However, within the present context the 
subset of those reparametrizations may be analysed which 
leaves the bosonic theory unchanged. Again the same number 
of arbitrary functions emerges for the different cases 
described in the paragraphs above. \\
A generic property of the fermionic extensions obtained in 
our analysis was the appearance of ``obstructions''. The 
first type of those consisted in singular functions of the 
bosonic variables $\phi$ and $Y$, multiplying the fermionic 
parts of a supergravity action, when no such singularities 
were present in the bosonic part. But even in the absence of 
such additional singularities, the relation of the original 
potential to some prepotential dictated by the 
corresponding supergravity theory, either led to a 
restriction of the range of $\phi$ and/or $Y$ as given by 
the original bosonic one, or even altogether prevented any 
extension of the latter. Remarkably, a known 2d supergravity 
model like the one of Howe \cite{Howe:1979ia} which 
originally had been  constructed 
with the full machinery of the superfield technique, escapes 
such obstructions. There, in our language, the PSM potential 
$v = -2 \lambda^2 
\phi^3$ permits an expansion in terms of the prepotential $u 
(\phi)$ through $v = -du^2/d\phi$. An example where 
obstructions seem to be inevitable is the KV-model \cite{Katanaev:1986wk} 
with quadratic bosonic torsion. \\
The hope that a link could be found between the possibility 
of reducing the arbitariness of extensions referred to above, and the 
absence of such obstructions, did not materialize. We could 
give several counter examples, including different singular 
and nonsingular extensions of SRG. \\
Another very important point concerns the ``triviality'', 
proved earlier by one of us \cite{Strobl:1999zz}. It was based upon the 
observation that locally a formulation of the dynamics in 
terms of Darboux  coordinates allows to elevate the 
infinitesimal transformations (\ref{gPSM-symms}) to finite ones. 
Then the  
latter may be used to gauge the fermionic fields to zero. 
Providing now the explicit form of those Darboux coordinates 
in the explicit solution of a generic model we also give 
additional support to the original argument of \cite{Strobl:1999zz}. 
However, the appearance of the obstructions and the ensuing 
singular factors in the transition to the Darboux 
coordinates may introduce a new aspect. When those new 
singularities appear at isolated points without restriction 
of the range for the original bosonic field variable, they 
may be interpreted and discarded much like coordinate singularities. 
Another way to circumvent this problem in the presence of 
restrictions to the range and thus to 
retain triviality is to 
allow a continuation of our (real) theory to complex 
variables.  This triviality disappears anyhow, when 
interactions with additional matter fields are introduced, 
obeying the same symmetry as given by the gPSM-theory. An 
example for this has been proposed already in ref.\  
\cite{Izquierdo:1998hg}. \\
In order to eliminate the arbitrariness of superdilaton 
extensions the only viable argument seems to consist in 
starting from a supergravity theory in higher dimensions 
(e.g.\ $D=4$) and to reduce it (spherically or  toroidally)  
 to a $D=2$ effective theory. However, the Killing spinors 
needed in that case must be Dirac spinors, requiring the 
generalization of the work \cite{EKS,Ertl:Diss} 
described here to (at least) 
$N=2$, where, however, the same technique of gPSM-s can  
be applied. \\[1.5cm]
\noindent
{\large\bf Acknowledgments}
\\[0.5cm]
This research has been supported by Austrian Science 
Foundation (FWF), project P-12815-TPH and P-14650-TPH.

\end{document}